\documentclass[aps,pre,floats,twocolumn,superscriptaddress,showpacs]{revtex4}

\usepackage{epsfig}
\usepackage{latexsym,amsmath}

\newcommand{\avk}{\langle k\rangle}
\newcommand{\fluck}{\langle k^2\rangle}

\begin{document}

\title{Diffusion-annihilation processes in complex networks}

\author{Michele Catanzaro}
\affiliation{Departament de F\'\i sica i Enginyeria Nuclear, Universitat
  Polit\`ecnica de Catalunya, Campus Nord B4, 08034 Barcelona, Spain}

\author{Mari\'an Bogu\~n\'a}
\affiliation{Departament de F\'\i sica Fonamental, Universitat de
  Barcelona, Av. Diagonal 647, 08028 Barcelona, Spain}

\author{Romualdo Pastor-Satorras}
\affiliation{Departament de F\'\i sica i Enginyeria Nuclear, Universitat
  Polit\`ecnica de Catalunya, Campus Nord B4, 08034 Barcelona, Spain}

\date{\today}

\begin{abstract}
  We present a detailed analytical study of the $A+A\to\emptyset$
  diffusion-annihilation process in complex networks. By means of
  microscopic arguments, we derive a set of rate equations for the
  density of $A$ particles in vertices of a given degree, valid for
  any generic degree distribution, and which we solve for uncorrelated
  networks. For homogeneous networks (with bounded fluctuations), we
  recover the standard mean-field solution, i.e. a particle density
  decreasing as the inverse of time. For heterogeneous (scale-free
  networks) in the infinite network size limit, we obtain instead a
  density decreasing as a power-law, with an exponent depending on the
  degree distribution. We also analyze the role of finite size
  effects, showing that any finite scale-free network leads to the
  mean-field behavior, with a prefactor depending on the network size.
  We check our analytical predictions with extensive numerical
  simulations on homogeneous networks with Poisson degree distribution
  and scale-free networks with different degree exponents.
\end{abstract}


\maketitle

\section{Introduction}
\label{sec:introduction}

In the last few years, complex networks have become a new paradigm for
complexity. The merging of graph theory together with new and
classical statistical physics tools has lead to the development of a
modern theory of complex network \cite{barabasi02,mendesbook}, that has
found fruitful applications in domains as divers as technology (the
physical Internet \cite{romuvespibook}, the World-Wide Web
\cite{hubbook}, power grids \cite{sixdegrees}), biology
(protein-protein interaction networks \cite{wagner01}, metabolic
networks \cite{Jeong00}, foodwebs \cite{montoya02,DiegoGuido}), social
sciences (sexual contact networks \cite{amaral01}, friendship networks
\cite{amaral}, scientific collaboration networks
\cite{newman01a,schubert}), etc.

The statistical analysis of many ``real-world'' networks has shown
that most of these systems seem to share some typical features, the
most relevant of them being the small-world property \cite{watts98}
and a large connectivity heterogeneity, reflected in the presence of a
scale-free degree distribution \cite{barab99}. The small-world
property refers to the fact that, in real networks, the hop distance
between two randomly chosen elements of the system is very small if
compared to the total number of elements. More precisely, if $\langle \ell \rangle $
is the average distance between two elements (or vertices), measured
as the smallest number of connections (or edges) between any vertices,
and $N$ is the system size (number of vertices) then, usually $\langle \ell \rangle$
increases logarithmically or slower with $N$. On the other hand,
scale-free networks are characterized by a degree distribution $P(k)$,
defined as the probability that a randomly selected vertex is
connected to $k$ other vertices (has degree $k$), that decreases as a
power-law,
\begin{equation}
  P(k) \sim k^{-\gamma},
\end{equation}
where $\gamma$ is a characteristic degree exponent, usually in the range $2
< \gamma < 3$. For these values of the degree exponent, the fluctuations in
the degree distribution, measured by the second moment $\langle k^2 \rangle$,
diverge in the infinite network size limit, $N\to\infty$, giving rise a very
heterogeneous connectivity lacking any characteristic degree scale.
This behavior is in opposition to the more classical homogeneous
networks \cite{erdos59}, which have a degree distribution decaying
exponentially or faster and exhibit bounded degree fluctuations.

Given that complex networks are widespread in nature, it becomes a
quite interesting issue the characterization of the effects that their
complex topology can have on dynamical processes taking place on top
of these systems. For example, it has been shown that heterogeneous
networks are remarkably weak in front of targeted attacks, aimed at
destroying the most connected vertices \cite{havlin01,newman00}, as
well as to the propagation of infective agents \cite{pv01a,lloyd01}.
These properties, which are mainly due to the critical interplay
between topology and dynamics in heterogeneous networks, are otherwise
absent in their homogeneous counterparts.

Epidemic processes, chemical reactions, and many other dynamic
processes, can all be modeled in terms of reaction-diffusion processes
\cite{vankampen}.  These are dynamic systems that involve particles of
different ``species'' ($A_i$, $i=1,\ldots n$) that diffuse stochastically
and interact among them following a fixed set of reaction rules.
One's interest is usually focused on the time evolution and steady
states of the densities of the different species $\rho_{A_i}(t)$, and the
possible presence of critical phase transitions \cite{marro99}. While
much is known about the behavior of reaction-diffusion processes on
regular homogeneous lattices, the situation is not so well established
in what respects the possible effects that a heterogeneous
connectivity structure can have on them.  At this respect, it is
interesting the work presented in Ref.~\cite{originalA+A}, in which a
numerical simulation analysis of the diffusion-annihilation process
$A+A\to\emptyset$ \cite{latticeA+A}, was performed on scale-free networks.  In
this reaction-diffusion process, particles of a single species $A$
diffuse on the vertices of a network and annihilate upon contact (when
two $A$ particles fall on the same vertex). In regular lattices of
Euclidean dimension $d$, it is well known that the local density of
$A$ particles, $\rho(x, t)$, is ruled by a Langevin equation
\cite{theoryA+A},
\begin{equation}
  \frac{ \partial \rho(x,t)}{\partial t} = D \nabla^2 \rho(x,t) - 2 \lambda \rho(x,t)^2 + \rho(x,t)
  \eta(x,t), 
  \label{eq:17}
\end{equation}
where $\eta(x,t)$ is an uncorrelated Gaussian noise. Dynamical
renormalization group arguments allow to show that the average
density of $A$ particles, $\rho(t) = \langle \rho(x,t)\rangle$, behaves in the large
time limit as
\begin{equation}
  \frac{1}{\rho(t)}-\frac{1}{\rho_0} \sim t^\alpha,
  \label{eq:16}
\end{equation}
where $\rho_0$ is the initial particle density, and the exponent $\alpha$
takes the values $\alpha = d / d_c$ for $d\leq d_c$ and $\alpha=1$ for $d>d_c$,
where $d_c=2$ is the critical dimension of this process. For $d> d_c$
one thus recovers the mean-field solution, obtained from
Eq.~(\ref{eq:17}) by setting the diffusion coefficient $D$ and the
noise term $\eta(x,t)$ equal to zero.

The numerical simulations of the $A+A\to\emptyset$ diffusion-annihilation
process reported in Ref.~\cite{originalA+A}, performed in scale-free
networks with general degree exponent $\gamma$, generated using the
configuration model \cite{bekessi72,benderoriginal,molloy95,molloy98},
led the authors to conclude that the behavior in time of the average
density of $A$ particles can be approximated by Eq.~(\ref{eq:16}),
where the asymptotic exponent $\alpha$ is a decreasing function of $\gamma$ and
is surprisingly larger than $1$ for $\gamma<3$.  The authors attributed
this effect to the small-world nature of the networks, and to the
existence of hubs (vertices with a large number of connections).

In spite of the potential interest of this result, no theoretical
arguments have been proposed so far to back up the numerical
conclusions reached in \cite{originalA+A}. In this paper we tackle
this task, by developing a mean-field analysis of the $A+A\to\emptyset$ process.
This analysis, made in the continuous $k$ approximation and inspired
in previous works made for epidemic spreading
\cite{pv01a,pv01b,marianproc}, results in a set of differential
equations for the density of $A$ particles in the vertices of degree
$k$, which are valid for networks with arbitrary degree distribution
$P(k)$ and two vertex correlations \cite{assortative}, determined by
the conditional probability $P(k'|k)$ that a vertex of degree $k$ is
connected to a vertex of degree $k'$ \cite{alexei,alexei02}.  The
solution of these equations for the particular case of uncorrelated
networks (in which the conditional probability $P(k'|k)$ is
independent of $k$) shows that, while homogeneous networks display a
pure mean-field behavior with exponent $\alpha=1$, scale-free networks with
$\gamma<3$ in the infinite size limit exhibit instead an exponent depending
on the properties of the network, i.e. $\alpha=1/(\gamma-2)$. Remarkably, this
solution in the infinite size limit shows a crossover for any finite
network to a linear behavior $1/ \rho(t) \sim t$, with a slope depending on
the network size. Our analytical results are confirmed by means of
large scale numerical simulations for both homogeneous and
heterogeneous networks.

We have organized the present paper as follows: In
Sec.~\ref{sec:a-+-o} we derive, from microscopic considerations, the
mean-field differential equations for the $A+A\to\emptyset$
diffusion-annihilation process in complex random networks with
arbitrary degree distribution and two vertex correlations, quantified
by means of $P(k)$ and $P(k'|k)$, respectively. We consider the case
of absence of correlations, finding the density of $A$ particles for
general homogeneous networks.  Sec.~\ref{sec:scale-free-networks} is
devoted to explicit results for scale-free networks, both in the
infinite size limit and for finite size networks.  In
Sec.~\ref{sec:numer-simul}, our analytic results are compared with
extensive numerical simulations of the diffusion-annihilation process
running on top of homogeneous and heterogeneous (scale-free) networks.
Finally, our conclusions are presented in Sec.~\ref{sec:conclusions}.

\section{The $\mathbf{A + A \to\emptyset}$ reaction in complex networks}
\label{sec:a-+-o}

Let us consider the diffusion-annihilation process $A + A \to\emptyset$ on a
complex network of size $N$ which is fully defined by the adjacency
matrix $a_{ij}$, that takes the values $a_{ij}=1$ if vertices $i$ and
$j$ are connected by an edge, and $0$ otherwise.  From a statistical
point of view, the network can also be characterized by its degree
distribution $P(k)$ and its degree correlations, given by the
conditional probability $P(k'|k)$.  Each vertex in the network can
host at most one $A$ particle, and the dynamics of the process is
defined as follows: Each particle jumps at a certain rate $\lambda$ to a
randomly chosen nearest neighbor. If it is empty, the particle fills
it, leaving the first vertex empty. If the nearest neighbor is
occupied, the two particles annihilate, leaving both vertices empty.

In order to study analytically this process in a general complex
network, in which vertices can show large degree fluctuations, we are
forced to consider the partial densities $\rho_k(t)$, representing the
density of $A$ particles in vertices of degree $k$, or, in other
works, the probability that a vertex of degree $k$ contains an $A$
particle at time $t$ \cite{pv01a,pv01b}.  From these partial
densities, the total density of $A$ particles is recovered from
\begin{equation}
  \rho(t) = \sum_k P(k) \rho_k(t).
  \label{eq:1}
\end{equation}

While it is possible to obtain a rate equation for the densities
$\rho_k(t)$ by means of intuitive arguments \cite{pv01a,marianproc}, in
the following we will pursue a more microscopical approach, which can
be generalized to tackle other kinds of problems. Let $n_i(t)$ be a
dichotomous random variable taking values $0$ or $1$ whenever vertex
$i$ is empty or occupied by an $A$ particle, respectively.  Using this
formulation, the state of the system at time $t$ is completely defined
by the state vector ${\bf n}(t)=\{n_1(t),n_2(t),\cdots,n_N(t)\}$. Assuming
that the time evolution of particles follows a Poisson process
\cite{vankampen}, the evolution of ${\bf n}(t)$ after a time increment
$dt$ can be expressed as
\begin{equation}
  n_i(t+dt)=n_i(t)\eta(dt)+[1-n_i(t)]\xi(dt),
  \label{evolution}
\end{equation}
where $\eta(dt)$ and $\xi(dt)$ are dichotomous random variables taking values
\begin{equation}
\eta(dt)=
\left\{
\begin{array}{cl}
0 & \mbox{with probability } \lambda dt \left[1+\displaystyle{\sum_j \frac{
      a_{ij} n_j(t)}{k_j}} \right] \\[0.5cm]  
1 & \mbox{otherwise}
\end{array}
\right.,
\end{equation}
and
\begin{equation}
\xi(dt)=
\left\{
\begin{array}{cl}
1 & \mbox{with probability } \lambda dt \displaystyle{\sum_j \frac{a_{ij}
    n_j(t)}{k_j}} \\[0.5cm] 
0 & \mbox{otherwise}
\end{array}
\right.
,
\end{equation}
where $a_{ij}$ is the adjacency matrix and $\lambda$ is the jumping rate
that, without loss of generality, we set equal to $1$. The first term
in Eq.~(\ref{evolution}) stands for an event in which vertex $i$ is
occupied by a particle and, during the time interval $(t,t+dt)$, it
becomes empty, either because the particle in it decides to move to
another vertex or because a particle in a nearest neighbor of $i$
decides to jump to $i$, annihilating thus both particles. The second
term corresponds to the case in which vertex $i$ is empty and a
particle in a neighbor vertex of $i$ decides to move to that vertex
\footnote{Notice that the random variables $\eta(dt)$ and $\xi(dt)$ are not
  independent, since both involve some common random movements. This
  fact, however, does not affect our further development.}.  Taking
the average of Eq.~(\ref{evolution}), we obtain
\begin{eqnarray}
  \lefteqn{\langle n_i(t+dt) | {\bf n}(t)\rangle=} && \nonumber \\  
  &=& n_i(t)-\left[n_i(t)-(1-2n_i(t))\sum_j \frac{a_{ij} n_j(t)}{k_j}
  \right]  dt,
\label{evolution_averaged} 
\end{eqnarray}
equation that describes the average evolution of the system,
conditioned to the knowledge of its state at the previous time step.
Then, after multiplying Eq.~(\ref{evolution_averaged}) by the
probability to find the system at state ${\bf n}$ at time $t$, and
summing for all possible configurations, we are led to
\begin{equation}
  \frac{d \rho_i(t)}{dt}=-\rho_i(t)+\sum_j a_{ij}\frac{1}{k_j}[\rho_j(t)-2\rho_{ij}(t)],
\label{rho_i}
\end{equation}
where we have introduced the notation $\rho_i(t) \equiv \langle n_i(t) \rangle$ and
$\rho_{ij}(t) \equiv \langle n_i(t)n_j(t)\rangle$. 

The derivation presented so far is exact. To proceed further, we
assume that vertices with the same degree are statistically equivalent
\cite{marianproc}. That is,
\begin{eqnarray}
  \rho_i(t) &\equiv& \rho_k(t) \qquad \forall i \in \mathcal{V}(k), \\
  \rho_{ij}(t)  &\equiv&  \rho_{kk'}(t) \qquad \forall i \in \mathcal{V}(k), \; j \in \mathcal{V}(k'),
\end{eqnarray}
where $\mathcal{V}(k)$ is the set of vertices of degree $k$.  Thus, by
summing Eq.~(\ref{rho_i}) for all vertices of degree $k$ and dividing by
the number of vertices with this degree, $N_k$, we can write, after
some formal manipulations,
\begin{equation}
\frac{d \rho_k(t)}{dt}=-\rho_k(t)+\sum_{k'}
\frac{\rho_{k'}(t)-2\rho_{kk'}(t)}{k'}\frac{1}{N_k}\sum_{i \in
  \mathcal{V}(k)}\sum_{j \in \mathcal{V}(k')}a_{ij}.
\end{equation} 
If all the vertices with the same degree are statistically equivalent,
we can set \cite{marianproc}
\begin{equation}
  \frac{1}{N_k}\sum_{i \in \mathcal{V}(k)}\sum_{j \in \mathcal{V}(k')}a_{ij}=kP(k'|k).
\end{equation}
Finally, by assuming the mean-field approximation $\rho_{kk'}(t) \simeq
\rho_k(t)\rho_{k'}(t)$, the rate equation for the density $\rho_k(t)$ can be
written as
\begin{equation}
  \frac{d \rho_k(t)}{d t} = - \rho_k(t) + k [ 1 - 2 \rho_k(t)] \sum_{k'}
  \frac{P(k' | k)}{k'}   \rho_{k'}(t).
  \label{eq:2}
\end{equation}

In the case of networks with general degree correlations, the solution
of Eq.~(\ref{eq:2}) depends on the nature of the conditional
probability $P(k' | k)$ and can be a rather demanding task
\cite{marian1}. Therefore, in the rest of this paper we will restrict
ourselves to the case of uncorrelated networks, in which the
conditional probability takes the simple form $P(k' | k) = k' P(k') /
\avk$ \cite{marianproc}. For this class of networks, the rate equation
Eq.~(\ref{eq:2}) is simplified to the form
\begin{equation}
   \frac{d \rho_k(t)}{d t} =  - \rho_k(t) + \frac{k}{\avk} [ 1 - 2 \rho_k(t)]
   \rho(t),
   \label{eq:3}
\end{equation}
where $ \rho(t)$ is the total density of $A$ particles. We can obtain a
differential equation for this last quantity by multiplying
Eq.~(\ref{eq:3}) by $P(k)$ and summing over $k$, namely,
\begin{equation}
   \frac{d \rho(t)}{d t} = -2 \rho(t) \Theta(t),
   \label{eq:4}
\end{equation}
where
\begin{equation}
  \Theta(t) = \frac{1}{\avk} \sum_k k P(k) \rho_k(t).
  \label{eq:5}
\end{equation}

In order to solve Eq.~(\ref{eq:4}), we perform a \textit{quasi-static}
approximation in the rate equation Eq.~(\ref{eq:3}). From the
mean-field solution of the $A+A\to\emptyset$ process, we expect $\rho(t)$ to be a
decreasing function with a power law-like behavior. In this case, for
large enough times, the time derivative of $\rho(t)$ will be much smaller
that the density proper. Extending this argument to the partial
densities $\rho_k(t)$, at large times we can neglect the left-hand-side
term in Eq.(\ref{eq:3}), and solve for $\rho_k(t)$ as a function of the
density, obtaining
\begin{equation}
  \rho_k(t) = \frac{k \rho(t) / \avk}{1 + 2 k \rho(t) / \avk}.
  \label{eq:6}
\end{equation}
Substituting this approximation into the expression for $\Theta(t)$, we get
\begin{equation}
  \Theta(t) = \frac{\rho(t)}{\avk^2} \sum_k \frac{k^2 P(k)}{1 + 2 k \rho(t) / \avk}.
\end{equation}
Inserting this last expression into Eq.~(\ref{eq:4}), we obtain as a
final equation for the density of $A$ particles
\begin{equation}
  \frac{d \rho(t)}{d t} = -2 \frac{\rho(t)^2}{\avk^2} \sum_k  \frac{k^2 P(k)}{1 + 2
    k \rho(t) / \avk}. 
  \label{eq:7}
\end{equation}

The solution of the approximate equation Eq.~(\ref{eq:7}) will depend
on the particular form of the degree distribution. The task becomes,
however, quite simple for the class of homogeneous networks.  In this
case, the degree distribution decreases so quickly that all its
moments are finite. So, for small $\rho(t)$ we can perform a Taylor
expansion of the right-hand-side of Eq.~(\ref{eq:7}), obtaining at
lowest order
\begin{equation}
   \frac{d \rho(t)}{d t} \simeq  -2 \frac{\fluck}{\avk^2} \rho(t)^2  + 4 \frac{\langle
     k^3\rangle}{\avk^3} \rho(t)^3,
\end{equation}
whose solution at large times yields
\begin{equation}
  \frac{1}{\rho(t)} - \frac{1}{\rho_0} \simeq   \frac{2 \fluck}{\avk^2} t,
  \label{eq:9}
\end{equation}
being $\rho_0$ the initial density of $A$ particles.  This corresponds to
the pure mean-field linear behavior, with a finite prefactor depending
on the second moment of the degree distribution.

Eq.~(\ref{eq:6}) can help us to understand how this process will
operate in heterogeneous networks. Indeed, for any given time $t^*$,
the partial density of vertices with degree larger than $\langle k \rangle
/2\rho(t^*)$ is essentially constant up to time $t^*$, that is, $\rho_k(t)\simeq
1/2$ for $k>\langle k \rangle /2\rho(t^*)$ and $t<t^*$.  The reason is that vertices
with high degree (hubs) are more easily reached by particles than
those with small degree and, with high probability they will be
surrounded by some $A$ particle.  Then, as soon as one nearby particle
decides to move to the hub, both particles disappear and another
nearby particle will replace the original one in the hub, keeping then
the density constant, which is clearly set to $1/2$. Therefore, hubs
act as drains through which particles vanish.  Besides, this process
happens in a hierarchical way in time, that is, the partial density of
small degree vertices decreases first, whereas the most connected
vertices are the last to hold particles.  This reasoning allows us to
anticipate that the resulting dynamics will strongly depend on the
number of high degree vertices of the network and, consequently, on
the degree distribution.

\section{Scale-free networks}
\label{sec:scale-free-networks}

For heterogenous networks with a diverging second moment, as in the
case of scale-free networks, we have to consider more carefully the
solution of Eq.~(\ref{eq:7}). Let us focus in generalized uncorrelated
scale-free networks in the infinite size limit, which are completely
determined by the normalized degree distribution
\begin{equation}
  P(k)=(\gamma-1) m^{\gamma-1} k^{-\gamma},
\end{equation}
where $2< \gamma <3$ is the degree exponent, $m$ is the minimum degree in
the network, and we are approximating $k$ as a continuous variable.
The average degree is thus $\avk = m (\gamma-1)/(\gamma-2)$. For a scale-free
network in the limit $N\to\infty$, Eq.~(\ref{eq:7}) can be written as
\begin{equation}
\begin{array}{lcr}
 \displaystyle{  \frac{d \rho(t)}{d t}} &=& -2 \rho(t)^2  \displaystyle{
   \frac{(\gamma-1)m^{\gamma-1}}{\avk^2}  
   \int_m^\infty 
   \frac{k^{2-\gamma}}{1+\frac{2 k \rho(t)}{\avk} } dk }\\[0.6cm]
   &=& - \rho(t) F\left(1, \gamma-2, \gamma-1, -\frac{(\gamma-1)}{2(\gamma-2)\rho(t)}\right),
\end{array}
\end{equation}
where $F$ is the Gauss hypergeometric function
\cite{abramovitz}. Therefore, the density as function of time can be
implicitly expressed in terms of the integral
\begin{equation}
  t = \int_{\rho_0^{-1}}^{\rho(t)^{-1}} \frac{d z}{z F(1, \gamma-2, \gamma-1,
    -\frac{(\gamma-1)z}{2(\gamma-2)})}.
    \label{eq:8}
\end{equation}
For very large times and small densities, we can use the asymptotic
expansion of the Gauss hypergeometric function, $F(1, \gamma-2, \gamma-1, -z)
\sim z^{2-\gamma}$, $z\to\infty$, to obtain the scaling behavior of the density with
time,
\begin{equation}
  \frac{1}{\rho(t)} \sim t^{1/(\gamma-2)}.
  \label{eq:12}
\end{equation}
This same result can be derived in a more intuitive fashion starting
from the argument presented at the end of the previous Section.  If
hubs act as drains of particles, then the rate of change in the total
density of particles will be proportional to the density of hubs. More
precisely, identifying the relevant hubs at time $t$ as those vertices
with degree larger than $\langle k \rangle /2\rho(t)$, we have that
\begin{equation}
  \frac{d \rho(t)}{d t} \sim \int_{\frac{\langle k \rangle}{2\rho(t)}}^\infty P(k) d k
  \sim \rho(t)^{\gamma-1},
\end{equation}
from where we obtain the same trend given by Eq.~(\ref{eq:12}). As $\gamma$
decreases the probability of having high degree vertices as drains in
the network increases. From Eq.~(\ref{eq:12}) we see as well that when
this happens, the decrease in the density gets faster. Therefore, the
bigger is the degree of the hubs in the network, the faster is the
process of annihilation.

We can obtain an exact solution for $\rho(t)$ in scale-free networks with
degree exponent $\gamma=3$. In this case, we have $P(k) = 2 m^2 k^{-3}$ and
$\avk = 2m$. The differential equation for the particle density reads
now
\begin{eqnarray}
  \frac{d \rho(t)}{d t} &=& - \rho(t)^2 \int_m^\infty \frac{k^{-1}}{1+k \rho(t)/m} d k
  \nonumber \\
  &=&    \rho(t)^2\ln\left( 1+\frac{1}{\rho(t)}\right),
\end{eqnarray}
whose solution is
\begin{equation}
  t=\int_{\rho_0^{-1}}^{\rho(t)^{-1}} \frac{d z}{\ln(1+z)} =
  \mathrm{L_i}(1+\rho(t)^{-1}) - \mathrm{L_i}(1+\rho_0^{-1}),
  \label{eq:10}
\end{equation}
where $\mathrm{L_i}$ is the logarithmic integral function
\cite{abramovitz}. For small $\rho(t)$, we can exploit the asymptotic
expansion of the logarithmic integral function, $\mathrm{L_i}(z) \sim z/
\ln z$, $z\to\infty$, to obtain
\begin{equation}
  \frac{1}{\rho(t)} \sim t \ln t.
  \label{eq:11}
\end{equation}
That is, in the infinite network size limit, the particle density of a
scale-free network with $\gamma=3$ follows the same decay as the mean-field
solution, with a logarithmic correction.

The conclusion of this analysis is that the particle density in the
$A+A\to\emptyset$ process in a infinite size scale-free network with degree
exponent $\gamma$ follows the asymptotic form at large times
\begin{equation}
   \frac{1}{\rho(t)} \sim t^{\alpha(\gamma)} (\ln t)^{\beta(\gamma)},
\end{equation}
with characteristic exponents 
\begin{eqnarray}
  \alpha(\gamma) &=& \left\{ \begin{array}{ccc}
      1/(\gamma-2)  & \qquad & 2<\gamma<3\\
      1 & \qquad & \gamma \geq 3
    \end{array} \right., \label{eq:18}\\
  \beta(\gamma) &=& \left\{ \begin{array}{ccc}
      0  & \qquad & 2<\gamma<3\\
      1 & \qquad & \gamma = 3\\
      0 & \qquad & \gamma > 3
    \end{array} \right..
\end{eqnarray}
The results for $\gamma>3$ are a natural consequence of the lack of degree
fluctuations ($\langle k^2 \rangle < \infty$) in this kind of scale-free networks.

The analytical exponents determined above, however, are strongly
affected by finite size effects. Indeed, for a power-law degree
distribution, the largest weight in the sum in Eq.~(\ref{eq:7}) is
carried by the large $k$ values. If the network is composed by a
finite number of vertices, $N$, as it always happens in numerical
simulations, it has a cut-off or maximum degree $k_c(N)$, which is
usually a function of the network size \cite{dorogorev}. Thus, there
exists a cross-over time $t_c$, defined by
\begin{equation}
  \frac{2 k_c(N) \rho(t_c)}{\avk} \ll 1,
  \label{eq:13}
\end{equation}
such that, for $t>t_c$ the particle density is so small that we can
approximate
\begin{equation}
  \Theta(t) \approx \frac{\rho(t)}{\avk^2}\sum_k k^2 P(k) = \rho(t) \frac{\fluck}{\avk^2}.
\end{equation}
In this time regime, we will observe an effective linear behavior for
the inverse particle density,
\begin{equation}
  \frac{1}{\rho(t)} \sim \frac{2 \fluck}{\avk^2} t,
  \label{eq:14}
\end{equation}
whose effective slope depends of the network size as
\begin{equation}
  \fluck = \sum_{k=m}^{k_c(N)} k^2 P(k) \sim k_c(N)^{3-\gamma}.
  \label{eq:15}
\end{equation}
That is, in finite size scale-free networks, the density of $A$
particles at very large times has the same behavior as in homogeneous
networks, i.e.  $1/ \rho(t)$ is linear in $t$, but now with a slope that
depends on the network size $N$.  Thus, the exponents larger than one,
given by Eq.~(\ref{eq:18}), can only be numerically observed as
transients in very large networks.

\section{Numerical simulations}
\label{sec:numer-simul}

As a check for the analytic predictions developed in the previous
Sections, we have performed extensive numerical simulations of the
diffusion-annihilation process on top of different model networks,
both homogeneous and heterogeneous.  The simulations are implemented
using a sequential updating scheme \cite{marro99}. An initial fraction
$\rho_{0}$ of vertices in the networks is randomly chosen and occupied by
an $A$ particle.  At time $t$ in the simulation, a vertex is randomly
chosen among the $n$ vertices that host an $A$ particle at that time.
One of its neighbors is selected also at random.  If it is empty, the
particle moves and occupies it.  If it contains a particle, this is
annihilated with the one in the first vertex, becoming both vertices
empty, and the number of occupied vertices is updated $n \to n-2$.  In
both cases, time is updated as $t \to t+ 1/n$, where $n$ is the total
number of particles at the beginning of the simulation step. In all
the simulations presented in this Section, we set the initial particle
density $\rho_0=0.5$.

\subsection{Homogeneous networks}

The class of homogeneous networks refers to networks with a degree
distribution peaked at an average value $\avk$ and decaying
exponentially or faster for $k \gg \avk$ and $k \ll \avk$. Typical
examples of such networks are the Erd\"os and R\'enyi model \cite{erdos60}
and the small-world model proposed by Watts and Strogatz (WS)
\cite{watts98}. We will focus in the latter in order to perform
computer simulations. The WS model is defined as follows
\cite{watts98,alain}: The starting point is a ring with $N$ vertices,
in which every vertex is symmetrically connected to its $2m$ nearest
neighbors.  Then, for every vertex, each edge connected to a
clockwise neighbor is rewired to a randomly chosen vertex with
probability $p$, and preserved with probability $1-p$. This procedure
generates a graph with a degree distribution that decays faster than
exponentially for large $k$, and average degree $\avk = 2m$.  We will
consider the WS model with $p=1$, that is, in which all edges have
been rewired. In this limit, the degree distribution of the WS network
takes the form \cite{alain}
\begin{equation}
  P(k) = \frac{m^{k-m}}{(k-m)!} e^{-m}.
\end{equation}
Therefore, all its moments $\langle k^n\rangle$ are finite for any value of $n$,
and we should expect Eq.~(\ref{eq:9}) to provide a good approximation
for the dynamics of the $A+A\to\emptyset$ process for $\rho(t) \to 0$.

In order to check this fact, we have carried out large scale numerical
simulations of the $A+A\to\emptyset$ reaction on WS networks with $p=1$ and
minimum degree $m=2$, which corresponds to an average degree $\avk=6$.
Simulations were performed on graphs of size $N=10^6$, averaging over
$1000$ reaction processes over $10$ different realizations of the
random network.
\begin{figure}
  \epsfig{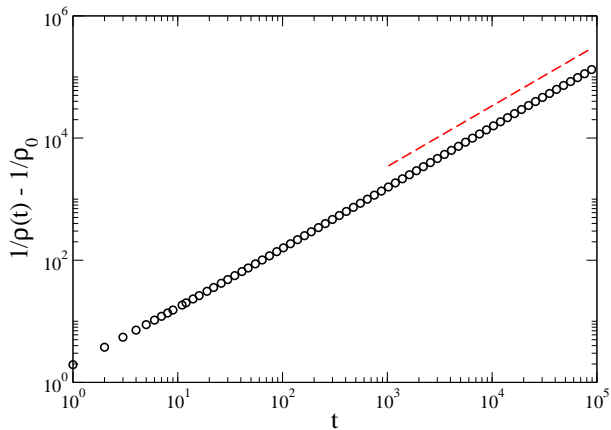}
  \caption{Inverse average particle density $\rho(t)$ as a function of time
    for the $A+A\to\emptyset$ process in WS networks. The dashed line
    corresponds to a behavior $1/ \rho(t) \sim t$.}
  \label{fig:wsdensity}
\end{figure}
In Fig.~\ref{fig:wsdensity} we show the total density of $A$ particles
as a function of time. As we can observe, the inverse particle density
follows quite precisely a linear behavior with time.  A linear fit for
the last two decades of $t$ values yields a slope $\approx 1.50$, which is
not too different from the predicted value in Eq.~(\ref{eq:9}), $2
\fluck / \avk^2 = 2.25$.

\subsection{Heterogeneous networks: The Barab\'asi-Albert model}

The Barab\'asi-Albert (BA) graph was introduced as the first growing
network model capable of producing as an emerging property a power-law
degree distribution \cite{barab99}. This model is based on the
preferential attachment paradigm, a rather intuitive mechanism in
which new individuals tend to develop more easily  connections with
already well connected individuals. The model is defined as follows:
We start from a small number $m_0$ of vertices, and at each time step,
a new vertex is introduced, with $m$ edges that are connected to old
vertices $i$ with probability
\begin{equation}
  \Pi(k_i) = \frac{k_i}{\sum_j k_j},  
\end{equation}
where $k_i$ is the degree of the $i$th vertex. After iterating this
procedure a large number of times, we obtain a network composed by $N$
vertices, minimum degree $m$, and average degree $\avk = 2m$, with a
degree distribution $P(k) = 2 m^3 k^{-3}$ and almost vanishing degree
correlations \cite{alexei02}. In the results presented here we
consider the parameters $m_0=5$ and $m=2$, corresponding to an average
degree $\avk=4$. Simulations were performed on networks of size
$N=10^5$ and $N=10^6$, averaging over $1000$ runs in $10$ different
network samples.

A first issue that we have explored in this network is the validity of
the quasi-stationary approximation made to obtain Eq.~(\ref{eq:6}). If
this approximation is valid, we should observe that the quantity
\begin{equation}
  k\left( \frac{1}{\rho_k(t)} -2 \right) = \frac{\avk}{\rho(t)}
\label{eq:26}
\end{equation}
is independent of $k$. 
\begin{figure}
  \epsfig{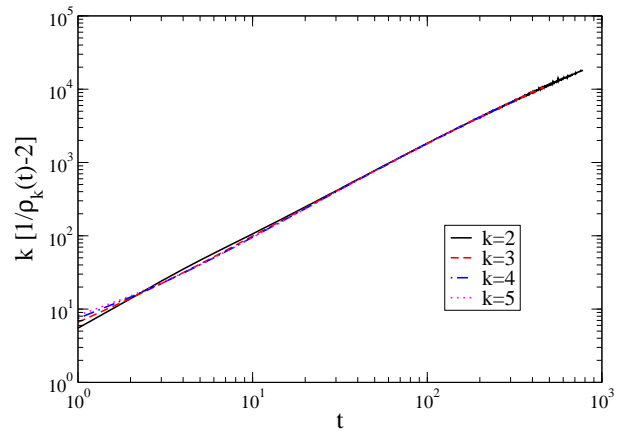}
  \caption{Check of Eq.~(\ref{eq:26}) in  BA networks. The networks
    have size $N=10^5$.}
  \label{fig:checkba}
\end{figure}
We have performed this check in Fig.~\ref{fig:checkba}. As we can
observe, the curves collapse quite nicely for $t$ larger that $t_0\simeq
10$. Therefore, we should expect that the results obtained from the
approximation in Eq.~(\ref{eq:7}) will be valid above this
characteristic time.

\begin{figure}
  \epsfig{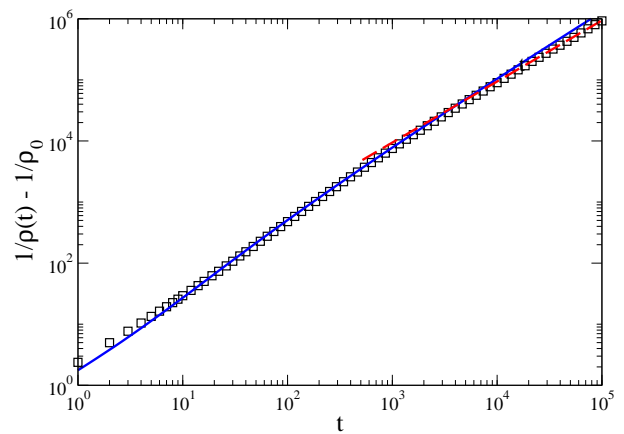}
  \caption{Inverse average particle density $\rho(t)$ as a function of time
    for the $A+A\to\emptyset$ process in BA networks. The full line is the
    analytic solution Eq.~(\ref{eq:10}); the dashed line corresponds
    to the finite size behavior $1/ \rho(t) \sim t$.}
\label{fig:BAdensity}
\end{figure}

In Fig.~\ref{fig:BAdensity} we represent the inverse particle density
as function of time. In this plot we observe that, except for very
early times, the initial time dependence is very accurately described
by the analytic solution Eq.~(\ref{eq:10}), which can be approximately
described by a linear behavior with a logarithmic correction, $1/ \rho(t)
\sim t \ln t$ (full line). At larger times, the function $1/ \rho(t)$
crosses over to a linear dependence (dashed line), showing the onset
of finite size effects.  The poor agreement at early times should be
attributed to the approximation in Eq.~(\ref{eq:7}), which is valid
only for times larger than $t_0 \simeq 10$.

\subsection{Heterogeneous networks: The uncorrelated configuration model}

The construction of uncorrelated scale-free networks with arbitrary
degree exponent is a nontrivial issue. Traditionally, this kind of
networks were generated using the configuration model (CM)
\cite{bekessi72,benderoriginal,molloy95,molloy98}, in which, starting
from a degree sequence extracted according to the desired degree
distribution, edges are randomly created between vertices, respecting
the preassigned degrees. It has been shown, however, that for
scale-free distributions with diverging second moment, and in the
absence of self-connections (a vertex joined to itself) and multiple
connections (two vertices connected by more than one edge), this
procedure generates in fact degree correlations
\cite{maslovcorr,originnewman}.  The origin of this phenomenon can be
traced back to the effects of the cut-off, or maximum expected degree,
$k_c(N)$, which must scale at most as $N^{1/2}$ in order to allow for
a full random uncorrelated edge assignment \cite{mariancutofss}.

To solve this question, it has been recently proposed the
\textit{uncorrelated configuration model} (UCM) \cite{ucmmodel}, that
is defined as follows:
\begin{enumerate}
\item Assign to each vertex $i$ in a set of $N$ initially disconnected
  vertices a degree $k_i$, extracted from the probability distribution
  $P(k)\sim k^{-\gamma}$, and subject to the constraints $m \leq k_i \leq N^{1/2}$
  and $\sum_i k_i$ even.
 \item Construct the network by randomly connecting the vertices with
   $\sum_i k_i /2$ edges, respecting the preassigned degrees and avoiding
   multiple and self-connections.
\end{enumerate}
Using this algorithm, it is possible to create scale-free networks
whose cut-off scales as $k_c(N) \sim N^{1/2}$ for any degree exponent
$\gamma$, and which are completely uncorrelated.  The results presented
below were obtained from simulations performed on networks generated
from the UCM algorithm with minimum degree $m=2$ and sizes ranging
from $N=10^5$ to $N=10^6$. Averages were performed over $1000$ runs in
$10$ different network.

\begin{figure}
  \epsfig{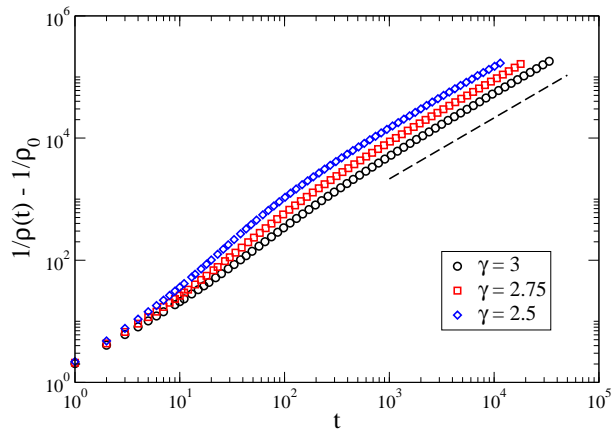}
  \caption{Inverse average particle density $\rho(t)$ as a function of time
    for the $A+A\to\emptyset $ process in uncorrelated scale-free networks with
    different degree exponents. The dashed line corresponds to the
    finite size behavior $1/ \rho(t) \sim t$.}
\label{fig:MRdensity}
\end{figure}

In Fig.~\ref{fig:MRdensity} we plot the inverse particle density from
computer simulations in networks with different degree exponent $\gamma$.
From this plot we observe that, at the initial time regime, the growth
of this function is faster for smaller values of $\gamma$, in agreement
with the theoretical prediction Eq.~(\ref{eq:12}). At larger times, on
the other hand, finite size affects take over, and we observe again a
linear regime (compare the slope with the reference dashed line in the
Figure). For $\gamma<3$, however, we cannot compare directly with the full
analytic prediction Eq.~(\ref{eq:12}), as we did for the BA model. The
reason is the following: From Eq.~(\ref{eq:13}), we have that the
density at the crossover time scales as
\begin{equation}
  \rho(t_c) \sim k_c(N)^{-1}.
\end{equation}
Then, from Eq.~(\ref{eq:12}), we can estimate
\begin{equation}
  t_c \sim k_c(N)^{\gamma-2}.
\end{equation}
That is, the crossover time $t_c$ diminishes with decreasing $\gamma$. It
turns out that, even for $\gamma=2.5$, $t_c$ is so small that we cannot
observe the transient regime $1/ \rho(t) \sim t^{1/(\gamma-2)}$, and only the
finite size effects, with its associated linear behavior, are
apparent.  System sizes larger than those available for our present
computer resources should make visible this crossover phenomena.

\begin{figure}[t]
\epsfig{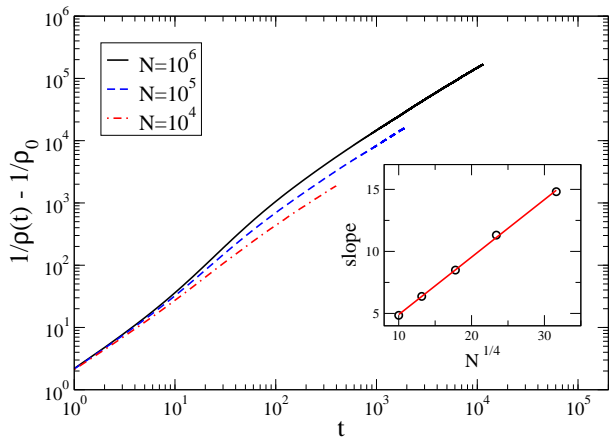}
\caption{Inverse average particle density $\rho(t)$ as a function of time
  for the $A+A\to\emptyset $ process in uncorrelated scale-free networks with
  degree exponent $\gamma=2.5$ and different network sizes. Inset: slope in
  the linear regime as a function of $N^{1/4}$. The good linear
  behavior validates the finite size solution in Eq.~(\ref{eq:14}).}
\label{fig:MRcheck}
\end{figure}

Finally, we have checked the slope dependence of the linear behavior
given by the finite size effects. According to Eqs.~(\ref{eq:14}) and
(\ref{eq:15}), we should observe, for a fixed degree exponent, a slope
that grows with the system size as $k_c(N)^{3-\gamma} \sim N^{(3-\gamma)/2}$, where
we have used the scaling of the cut-off given by the UCM algorithm.
We have analyzed this point in Fig.~\ref{fig:MRcheck}. For a degree
exponent $\gamma=2.5$, we observe that the curves for increasing values of
$N$ show a correspondingly increasing of the slope in the final linear
region. We have estimated this slope by performing a linear fitting to
the whole curve (inset in Fig.~\ref{fig:MRcheck}). This slope
increases approximately as $N^{1/4}$, in very good agreement with our
theoretical predictions.

It is noteworthy that these finite size effects, so remarkable in the
simulations presented in this Section, are however not evident in the
numerical work developed in Ref.~\cite{originalA+A}. The reason of
this fact could be that the simulations in \cite{originalA+A} were
performed in networks generated with the CM algorithm. These networks,
as we have discussed above, exhibit some degree of correlations
\cite{maslovcorr,originnewman,ucmmodel}, and therefore could be
outside of the regime of validity of the analytical calculations we
have presented here.

\section{Conclusions}
\label{sec:conclusions}

In this paper we have presented a detailed analytical study of the
$A+A\to\emptyset$ diffusion-annihilation process in complex networks
\cite{originalA+A}. For uncorrelated homogeneous networks with bounded
degree fluctuations, we recover a behavior compatible with the
standard mean-field solution of the process, that is, the inverse
particle density grows linearly with time, $1/ \rho(t) \sim t$, with a
constant prefactor. In the case of uncorrelated heterogeneous
networks, characterized by a scale-free degree distribution, we
observe instead that, in the infinite network size limit $N\to\infty$, the
inverse particle density increases as a power law with time, $1/ \rho(t)
\sim t^{\alpha(\gamma)}$, with an exponent larger than one and that depends on the
level of heterogeneity of the network, i.e. $\alpha(\gamma) = 1/(\gamma-2)$ for
$\gamma<3$. For $\gamma=3$, a linear growth with logarithmic corrections sets
in, while for $\gamma>3$ we recover again the mean-field solution typical
of homogeneous networks. In the case of scale-free networks, we have
also analyzed the effects of a finite network size in this dynamical
process.  For any value of $\gamma\leq3$, we have shown that, at large times,
the inverse particle density in a finite network crosses over to a
linear behavior, with a slope that is an increasing function of the
network size $N$, i.e. $1/\rho(t) \sim N^{(3-\gamma)/2} t$. In order to check our
results, we have performed extensive numerical simulations on both
homogeneous and heterogeneous networks, obtaining a convincing
evidence for our predictions.

An interesting conclusion that can be extracted of the present work is
the very strong incidence that finite size effects can have on
dynamical systems on top of scale-free networks. While these size
effects have been already discussed in the context of epidemic
spreading \cite{pvbrief}, the $A+A\to\emptyset$ process analyzed here provides a
further example, in which analytical results and numerical simulations
show a striking agreement.

\begin{acknowledgments}
  We thank A. Vespignani and F. van Wijland for helpful comments and
  discussions. This work has been partially supported by EC-FET Open
  Project No.  IST-2001-33555.  R.P.-S. acknowledges financial support
  from the Ministerio de Ciencia y Tecnolog\'\i a (Spain), and from the
  Departament d'Universitats, Recerca i Societat de la Informaci\'o,
  Generalitat de Catalunya (Spain). M. B. acknowledges financial
  support from the Ministerio de Ciencia y Tecnolog\'\i a through the
  Ram\'on y Cajal program. M. C. acknowledges financial support from
  Universitat Polit\`ecnica de Catalunya.
\end{acknowledgments}

\end{document}